# Effects of Local Atomic Order on the Pre-Edge Structure in the Ti *K* X-Ray Absorption Spectra of Perovskite $CaTi_{1-x}Zr_xO_3$

V. Krayzman,[a] I. Levin, J. C. Woicik, D. Yoder, and D. A. Fischer

Ceramics Division, National Institute of Standards and Technology, Gaithersburg, Maryland 20899, USA

**Abstract**

The effects of local B-cation (Ti, Zr) distribution and octahedral tilting on the pre-edge structure in the Ti X-ray absorption *K*-spectra of the $CaTi_{1-x}Zr_xO_3$ perovskite solid solutions were investigated. Experimental spectra for the disordered $CaTi_{1-x}Zr_xO_3$ samples revealed systematic variations of the pre-edge peak intensities with the *x*-values. Multiple-scattering calculations using 75-atom clusters $Ti(TiO_6)_{6-n}(ZrO_6)_nCa_8O_{24}$ were conducted to interpret these differences. The origin of the lowest unoccupied states in the conduction band of the $CaTi_{1-x}Zr_xO_3$ was determined from the analyses of X-ray absorption near-edge structure of the O *K*-edge. The calculations reproduced the experimental spectra and demonstrated that the differences in the intensities of certain pre-edge feature are dominated by the probability of finding a Zr atom in the first B-cation coordination sphere around the absorbing Ti. The pre-edge structure appeared to be sensitive to small changes in the value of this probability, so that the pre-edge intensities could be used effectively to compare the extent of local B-site order in perovskite solid solution samples having similar chemical composition but processed differently.

---

[a] Email: victor.krayzman@nist.gov,

77.84.Dy, 78.70.Dm

# 1. INTRODUCTION

Many complex metal oxides crystallizing with perovskite-like structures exhibit exploitable functional properties (i.e. dielectric, electronic, magnetic, *etc*.). Commonly, these properties are sensitive to slight variations in lattice distortions and chemical/displacive atomic ordering; therefore, understanding of structure-properties relations becomes critical for a development of new and optimization of the existing materials. Most of to-date research on structure-property relations in perovskites has focused on the long-range structural effects (e.g. cation ordering, octahedral tilting, cation displacements) which can be commonly deduced from the Bragg diffraction. Despite numerous examples of functional responses in perovskite-like oxides (especially the solid solutions) being dominated by the details of local atomic arrangements,[1,2] the understanding of structural behavior at this scale remains incomplete largely because of the difficulties associated with local structure measurements.

Local deviations from the average structure are manifested in the details of diffuse X-ray/neutron scattering and X-ray absorption fine structure (XAFS). Unlike typical diffraction data, structural information contained in XAFS is element-specific. Extended XAFS (EXAFS) is used commonly to access the atomic environment around the absorbing species. For perovskites based on 3*d* and 4*d* metal ions, a pre-edge XAFS in transition metal *K*-edge spectra also carries information about (i) local distortions of the oxygen coordination environment and (ii) distribution of the neighboring 3*d* and 4*d* cations.[3, 4, 5, 6] The pre-edge structure, therefore, is expected to be sensitive to the short-range B-site order and can be used to complement the EXAFS results; yet, no systematic studies of the pre-edge transition metal XAFS in terms of the local B-cation ordering have been reported. In the present study, we addressed this issue by analyzing the effects of B-cation distribution and octahedral tilting on the Ti *K*-pre-edge structure in the perovskite solid solutions $CaTi_{1-x}Zr_xO_3$. Additionally, we report a systematic analysis of the X-ray absorption near-edge structure (XANES) for the O *K*-edge in this system. The analyses of O *K*-XANES were used for identification of the lowest-energy conduction bands in $CaTi_{1-x}Zr_xO_3$ thus facilitating interpretation of a more complicated Ti *K*-pre-edge structure.

The $CaTi_{1-x}Zr_xO_3$ solid solutions were chosen for this study as they represent a common class of perovskite-like dielectrics which incorporate dissimilar-size cations (ionic radii $R_{Ti}$=0.604 Å, $R_{Zr}$=0.70 Å)[7] on the B-sites and exhibit tilting of the [$BO_6$] octahedra. Both end-compounds, $CaTiO_3$ and $CaZrO_3$, crystallize with similar room-temperature structures having orthorhombic *Pbnm* symmetry and lattice parameters $\sqrt{2}a_c\times\sqrt{2}a_c\times 2a_c$ ($a_c\approx 4$ Å is the lattice parameter of an ideal cubic perovskite).[8,9] The deviations from the ideal perovskite symmetry in both structures are caused by the $a^+b^-b^-$ [10] octahedral tilting; however, the degree of tilting is greater in $CaZrO_3$.

Recent analyses of local structures in the well-equilibrated $CaTi_{1-x}Zr_xO_3$ solid solutions using combined EXAFS and neutron scattering pair-distribution function methods revealed that both Zr-O and Ti-O distances in the solid solutions remain close to their respective values in the end-compounds.[11] Additionally, the analyses suggested a three-modal distribution of the octahedra tilting angles in the solid solutions corresponding to the Ti-O-Ti, Zr-O-Zr, and Ti-O-Zr links. Semi-quantitative analyses of the EXAFS and PDF data supported a near-random distribution of Zr and Ti over octahedral B-sites. However, accurate analyses of short-range B-site order using EXAFS were precluded by the strong correlations among the occupational probability and the octahedra tilting angles. In this contribution, we examined the sensitivity of a Ti *K*-pre-edge XAFS to the short-range B-site order in complex Ti-based perovskites.

## II. EXPERIMENTAL

The $CaTi_{1-x}Zr_xO_3$ powder samples used in this work were identical to those used previously in the EXAFS and neutron scattering studies; the details of sample preparation can be found in Ref. [11]. In short, samples with $x$=0, ¼, ½, ¾, 1 were prepared by conventional solid state synthesis and equilibrated by multiple heating (with intermediate grinding) at 1500 ºC for a total of 500 h. The equilibrium was inferred from the lack of change in the X-ray line broadening.

The Ti *K*- and O *K*-XANES were measured using the National Institute of Standards and Technology beamlines X23A2 and U7A, respectively, at the National Synchrotron Light Source. The Ti *K*-edge data were collected in transmission. For these measurements, the double crystal monochromator was operated with a pair of Si (311) crystals; a detuning was used to minimize the effect of harmonics. The O *K*-XANES were measured with partial electron yield utilizing a totally enclosed channeltron with a three grid high pass electron analyzer held at a bias of -50 V. The insulating powder samples were also charge neutralized by a low energy electron flood gun. The partial electron yield signal was normalized by the photo-yield from a freshly evaporated gold mesh placed in the incident beam before the sample to remove monochromator absorption features and beam instabilities.

## III. CALCULATION PROCEDURE

Theoretical Ti *K*- and O *K*-absorption spectra were calculated using a full multiple-scattering method implemented in the XKDQ computer code[3] that employs a muffin-tin (MT) cluster potential and includes both dipole and quadrupole parts of electron-photon interactions. Since all one-electron methods which rely on the MT model are approximate anyway, we adopted a semi-empirical method[3] for constructing the cluster potential. In this method, the potentials of all atoms in the cluster but the absorbing one were calculated from the electron densities of the free neutral atoms using a Hermann-Skillman procedure.[12] The corrections for atomic potentials $\Delta E_i$ were introduced to account for the overlap of atomic potentials, the Madelung potential, and the redistribution of electrons in the crystal. These corrections, assumed to be constant within each atomic sphere, were chosen to provide the best agreement with the experimental spectra. According to our calculations, the typical absolute values of $\Delta E_i$ were less than 5 eV. The same sets of $\Delta E_i$ were used for both Ti *K*- and O *K*-spectra. The $\alpha$ parameter in the X$\alpha$ exchange-correlation term for atomic electron density calculations was chosen according to the Schwartz prescription.[13] The exchange-correlation term in the calculations of the photoelectron scattering phase shifts and atomic matrix elements was determined using the Hermann-Skillman electron densities and $\alpha$=0.6. The electron scattering phase shifts, calculated according to this approach, were close to those calculated using the Hartree-Fock approximation.

The radii of non-overlapping MT spheres were chosen to minimize the potential discontinuities at the sphere boundaries.

XANES real-space multiple-scattering calculations that involve the “number of atom–orbital momentum” Green’s function representation enable unambiguous identification of (1) the mechanisms giving rise to various pre-edge peaks and (2) the molecular orbitals (MO) corresponding to the final states of the excited photoelectron. The latter is achieved either by suppressing a photoelectron scattering from selected atoms for a particular orbital momentum or by varying the potential shift and tracking the change in the peak intensities and positions.

The influence of a core-hole potential on the pre-edge structure as well as the correspondence between this structure and the local partial density of states in the crystal ground state have been discussed extensively. While the core-hole potential produces a relatively small effect on the O *K*-pre-edge,[14, 15] its effect on the metal *K*-pre-edge structure is dramatic.[3] According to the studies of Ti *K*-edge XANES in perovskites and rutile, neither a popular (Z+1) approximation nor self-consistent cluster calculations reproduce the experimental pre-edge structure. A reasonable agreement with the experimental data has been obtained by using the incomplete screening model[3] and full-potential simulations,[16] as well as by solving a Bethe-Salpiter equation.[17] In the present study, we employed the incomplete screening approach with the screening charge of 0.8 electron charges to account for the core-hole effect.

## IV. RESULTS AND DISCUSSION

### A. O *K*-XANES and the lowest-energy conduction bands in $CaTi_{1-x}Zr_xO_3$

O *K*-XANES in $ATiO_3$ (A=Ba, Sr) perovskites has been the subject of numerous experimental and theoretical investigations.[14, 15, 18, 19, 20, 21] The theoretical techniques used to calculate the O *K*-edge in perovskites included band-structure calculations,[14, 15, 21] the discrete-variational $X\alpha$ method,[18] and multiple-scattering calculations.[19, 20] Most theoretical studies focused on the O *K*-edge in $SrTiO_3$ which crystallizes with an ideal cubic perovskite structure. While the origins of various features in the O *K*-XANES of $SrTiO_3$ have been established over a decade ago, a reasonable qualitative agreement between the calculated and experimental spectra has not been achieved until very recently.[20] In contrast to $SrTiO_3$, few systematic studies of O *K*-XANES have been reported for other perovskite-like compounds and their solid solutions.[21] In the present paper, we used the lower energy part of the O *K*-XANES, which corresponds to the pre-edge region in the Ti *K*-XANES, for identifying the peaks associated with the Ti and Zr *d*-conduction bands .

O *K*-XANES for the $CaTi_{1-x}Zr_xO_3$ solid solutions are summarized in Fig. 1. The nature of individual peaks in the spectra of $CaTiO_3$ and $CaZrO_3$ was determined from the calculations by (1) suppressing the *d*-scattering from the neighboring B-cations (Ti or Zr) and (2) varying the corrections $\Delta E_i$ in the B-cation atomic spheres. The absence of *d*-scattering caused peaks A and B for $CaTiO_3$ to disappear completely, whereas a variation of the corrections $\Delta E_i$ altered the energies of these peaks; similar effects were observed for peak B in $CaZrO_3$. For $CaTiO_3$, the origins of peaks A, B, and C appeared to be similar to those of the corresponding peaks in $SrTiO_3$.[14-16, 18, 19] In particular, peak A is associated with the anti-bonding $t_{2g}$-type MO constructed mainly from the O 2*p* atomic orbitals (AO) and the 3*d* AO of the two neighboring Ti atoms. Peak B corresponds to the anti-bonding $e_g$-type MO of the same atoms. Peak A is much stronger than peak B because in both $CaTiO_3$ and $CaZrO_3$ the $t_{2g}$ band is considerably narrower than the $e_g$ band. Peak C, featuring a complicated shape, arises due to a hybridization of the O 2*p* AO and Ca 3*d* AO. The origin of peak B in $CaZrO_3$ is analogous to the origin of peak A in $CaTiO_3$; however, in $CaZrO_3$, this peak occurs at higher energy because the energy of the Zr 4*d* level is higher than that of Ti 3*d*. Furthermore, the Zr 4*d* AO are more delocalized than the Ti 3*d* AO, giving a larger $t_{2g}$-$e_g$ splitting in $CaZrO_3$. Peak C in $CaZrO_3$ includes contributions from the photoelectron excitations to both $e_g$-type O 2*p* - Zr 4*d* MO and O 2*p* - Ca 3*d* MO. This assignment of peaks A and B is entirely consistent with the dependence of their intensities on the Ti/Zr ratio in the solid solutions: The intensity of peak A in the $CaTi_{1-x}Zr_xO_3$ samples is approximately proportional to the Ti concentration, whereas the intensity of peak B is a linear combination of peak B intensities in the spectra of $CaTiO_3$ and $CaZrO_3$.

Thus, the two lowest-energy conduction bands in the $CaTi_{1-x}Zr_xO_3$ system are narrow and well-separated from the higher-energy bands. These two bands correspond (in the order of increasing energy) to the Ti $t_{2g}$ band and a superposition of the Ti $e_g$ and Zr $t_{2g}$ bands, respectively.

### B. Ti *K*-pre-edge structure: effects of short-range cation order and octahedral tilting

The Ti *K*-spectra for the $CaZr_{1-x}Ti_xO_3$ samples are summarized in Fig. 2. Detailed analyses of pre-edge features in the Ti *K*-spectra of several perovskite-like structures have been reported previously.[3] According to these studies, peak A is caused primarily by the quadrupole excitation of 1*s* electron to the $t_{2g}$ orbitals of the absorbing [$TiO_6$] octahedron, whereas peak B arises from the excitation of 1*s* electron to the $e_g$ orbitals of the same octahedron and contains both quadrupole and dipole (caused by the *p*-*d* mixing) contributions.

Despite Ti atoms in $CaTiO_3$ occupying centrosymmetric positions, thermal vibrations (u=0.1 Å) induce a *p*-*d* mixing[3] sufficient to account for the observed magnitude of peak B. The intensity of this peak remains nearly constant across the solid solutions, which indicates that the Ti atoms retain their centrosymmetric environments over the entire compositional range. The energies of both $t_{2g}$ and $e_g$ orbitals for the absorbing [$TiO_6$] octahedron exhibit downshifts under the influence of attractive potential of the core-hole; therefore, peaks A and B are often referred to as excitonic.

The next two features, C and D, in the Ti pre-edge of oxide structures based on [$TiO_6$] octahedra have been attributed to the dipole excitations of the Ti 1*s* electrons into the $t_{2g}$ and $e_g$ orbitals of the neighboring [$TiO_6$] octahedra, respectively.[3, 22] The intensity of peak C is known to depend strongly on both the conjugation of [$TiO_6$] octahedra (i.e. edge- or corner-sharing) and their distortions. For example, this peak, being very prominent in rutile[22] (edge-sharing), is barely detectable in tetragonal perovskite $PbTiO_3$, and absent completely in cubic perovskites $SrTiO_3$ and $EuTiO_3$.[3] These studies demonstrated that the effect of thermal vibrations on features C and D is negligible.[3]

Our multiple-scattering calculations reproduced closely peaks A, B, and D in the experimental spectra of $CaTiO_3$, whereas feature C in the calculated spectra appeared as a shoulder on the high-energy side of peak B. The calculations were conducted using 75- and 200-atom clusters (Fig. 3). The 75-atom cluster (Fig. 4) included only the central (absorbing) [$TiO_6$] octahedron, six neighboring [$TiO_6$] octahedra, and eight neighboring [$CaO_{12}$] polyhedra. The central Ti atoms in both clusters had to be displaced by 0.1 Å from the ideal positions to simulate the effect of thermal vibrations that give rise to peak B.[b] The assignment of peaks C and D was confirmed by suppressing the *d*-scattering from the neighboring Ti atoms in the calculations: the resulting spectra were similar to those presented in Fig. 3 but lacked features C and D; thus both features originate from the transitions to the orbitals of the neighboring octahedra. Furthermore, slight variations of $\Delta E_i$ for the MT potentials of the neighboring Ti atoms produced analogous changes in the energies of these two pre-edge features.

The calculations revealed that the complex structure above peak D is caused by the transitions to the Ca *d* bands. Similar to feature C, the agreement between the calculated and experimental data for this part of the spectrum is significantly worse than that obtained for peaks A, B, and D. Clearly, the low-energy features above the excitonic peaks are sensitive to the exact shape of potential between the atoms, and, therefore, the full-potential calculations are needed for accurate reproduction of these parts of the spectra. Nevertheless, calculations performed under the MT potential approximation succeed in reproducing variations of the spectra due to octahedral tilting and local B-cation order, as discussed below.

[b] Since no significant differences was observed between the spectra calculated for the 75- and 200-atom clusters, the 75-atom cluster was used for the pre-edge structure calculations in the solid solutions.

Unlike peaks A and B, both features C and D display a pronounced dependence on Zr content. We isolated the compositional dependence of these features by subtracting the spectra of $CaTi_{3/4-x}Zr_{1/4+x}O_3$ from those of $CaTi_{1-x}Zr_xO_3$ with x=0, ¼, ½. The resulting difference spectra yield two peaks between 0 eV and 6 eV (Fig. 5). These two difference peaks, which occur at the energies corresponding to features C and D, reflect variations in the magnitude of dipole excitations of photoelectrons to the $t_{2g}$ and $e_g$ orbitals of the neighboring [$TiO_6$] octahedra, respectively. Our analyses of the lowest energy conduction bands from the O *K*-XANES suggest close energies for the Zr $t_{2g}$ and Ti $e_g$ bands; therefore, dipole excitations of the Ti 1*s* electrons to the $t_{2g}$ orbitals of the neighboring [$ZrO_6$] octahedra also contribute to the higher-energy difference peak in the solid solutions.

The observed variations of features C and D across the solid solutions could be caused by the two structural effects: (1) change in the probability, $p_{Zr}$, of finding a Zr atom on the six neighboring B-sites around the absorbing Ti and (2) change in the octahedra tilting angles which increase monotonically[11] with increasing Zr-content. The probability $p_{Zr}$ is directly related to the Ti/Zr short-range order parameter $\vartheta=1-p_{Zr}/x$.[23] The extent of these two effects on the Ti *K*-pre-edge structure was examined using a series of clusters, as described below.

A variation of the octahedra tilting angle affects the intensities of the pre-edge peaks via (i) changes in the hybridization of the *p* states on the absorbing Ti atom with the $e_g$ and $t_{2g}$ orbitals of the neighboring octahedra and (ii) Ca displacements modifying the narrow Ca *d*-bands. The Ti *K*-spectra were calculated for the 75-atom clusters of $CaTiO_3$ with the tilting angles varying from 10º to 20º, as encountered across the $CaTi_{1-x}Zr_xO_3$ solid solutions[11]. The model $CaTiO_3$ structures with different tilting angles were generated using the SPuDS software[24] that assumes rigid $[BO_6]$ octahedra and optimizes the atomic positions according to the bond valence sum criteria.

According to our calculations, the intensities of peaks that originate from the transitions to the $t_{2g}$ (i.e. feature C) and $e_g$ (peak D) orbitals of the neighboring octahedra increase and decrease, respectively, with the tilting angle increasing. These results explain lack of the C-like features in the spectra of cubic perovskites $SrTiO_3$ and $EuTiO_3$. However, the dependence of intensities on the tilting angle is in contrast to the behavior of peak C in the experimental spectra of the $CaTi_{1-x}Zr_xO_3$ samples and, therefore, cannot account for the observed variation of the pre-edge peaks intensities. The fluctuations of B-O-B angles due to thermal vibrations are significantly smaller than to the average tilt angles and, therefore, can be neglected.

The effect of Zr-content in the first B-cation coordination shell of the absorbing Ti atom on the pre-edge structure was evaluated using the 75-atom clusters featuring different fractions and distributions of $[TiO_6]$ and $[ZrO_6]$ octahedra. The solid solution clusters were generated by replacing the $[TiO_6]$ octahedra in the cluster built for $CaTiO_3$ with the $[ZrO_6]$ octahedra extracted from the $CaZrO_3$ structure. The $[ZrO_6]$ octahedra were attached to the central $[TiO_6]$ octahedron so that they maintained the same tilting angle as in their original location in the $CaZrO_3$ structure; that is, the resulting Ti-O-Zr angles were close to the average of the Ti-O-Ti and Zr-O-Zr angles in the respective end-compounds. Thus, the solid solution clusters accounted for a variation of the local tilting angles with the change in the nature of the neighboring B-cations.

Fig. 6 presents a series of difference spectra obtained by subtracting the spectra of the solid solution clusters $Ti(TiO_6)_{6-n}(ZrO_6)_nCa_8O_{24}$ from the spectra of $Ti(TiO_6)_6Ca_8O_{24}$. According to the calculations, the intensities of the two difference peaks at ≈0.8 eV and ≈3.5 eV increase with the number of $[ZrO_6]$ octahedra increasing. This trend is expected for the lower energy peak which is associated with transitions to the $[TiO_6]$ states. The behavior of the higher energy peak, which results from transitions to the states of the neighboring $[TiO_6]$ and $[ZrO_6]$ octahedra reflects a greater intensity of excitations to the $e_g$ states of the $[TiO_6]$ as compared to the excitations to the $t_{2g}$ states of $[ZrO_6]$ octahedra. In addition to this main trend, significant differences are observed among the spectra calculated for the $Ti(TiO_6)_{6-n}(ZrO_6)_nCa_8O_{24}$ clusters having fixed $n$-value but distinct arrangements of Zr and Ti. Our calculations prove that the pre-edge structure is sensitive not just to the number of $[ZrO_6]$ octahedra around the absorbing Ti atom but also to the spatial configurations of Ti and Zr.

For the *Pbnm* symmetry, the number of non-equivalent configurations in the $Ti(TiO_6)_{6-n}(ZrO_6)_nCa_8O_{24}$ clusters is 2 for $n$=1, 5 and 4 for $n$=2, 3, 4. Let us assume that for a given $n$, the probability of finding a particular $j^{th}$ configuration of the $[TiO_6]$ and $[ZrO_6]$ octahedra depends only on the statistical weight of this configuration, $w_j(n)$. In this case, the absorption spectra for the solid solutions $I_s(p_{Zr})$ are determined exclusively by the $p_{Zr}$ value:

$$I_s(p_{Zr}) = \sum_{n=0}^{6} C_6^n p_{Zr}^n (1-p_{Zr})^{6-n} < I(n) > \quad (1)$$

where $C_6^n$ are the binomial coefficients and the $\langle I(n)\rangle$ is the average absorption for a cluster with $n$ [$ZrO_6$] octahedra which can be calculated as:

$$< I(n) >= \sum_j \frac{w_j(n) I_j(n)}{\sum_j w_j(n)} \qquad (2)$$

The differences $I_s$ ($p_{Zr}$)-$I_s$ ($p_{Zr}$+0.1) for $p_{Zr}$=0, 0.1, 0.2…, and 0.9 are shown in Fig. 7. These results confirm that the intensities of features C and D in the solid solutions are determined primarily by the number of Zr atoms in the first B-cation coordination shell of the absorbing Ti atom. The difference intensities form a nearly-symmetrical function of $p_{Zr}$ having the minimum at the $p_{Zr}$=½ and increasing toward the end-compositions. Thus, for a completely random B-cation distribution ($\vartheta$=0), the following relation is expected: $½(I(p_{Zr})+I(1-p_{Zr}))\cong I(½)$. The experimental spectra for the $x$-values of ¼, ½, and ¾ obey this relation very closely (Fig. 8) being consistent with a near-random distribution of Ti and Zr, as suggested by the EXAFS and PDF analyses.[11]

The theoretical difference spectra calculated for the pairs of $p_{Zr}$ (¾-½), (½-¼), (¼-0) (Fig. 9) reproduce those calculated from the experimental spectra for the analogous compositions. This good agreement, while providing a further support for a near-random distribution of Zr and Ti in the presently studied solid solutions, indicates that our calculations capture the main physics underlying the Ti pre-edge structure in perovskites.

The pre-edge structure calculated for $\vartheta$=-0.1 (ordering) and $\vartheta$=+0.1 (clustering) for the compositions $x$=¼, ½, ¾ reveal significant systematic variation of the pre-edge intensities with $\vartheta$ (Fig. 10). Thus, the Ti *K*-pre-edge structure can be used as a sensitive tool for a semi-quantitative comparison of the short-range B-cation order parameters in the samples having the same chemical composition but processed differently. For example, this technique can be attractive for studying the effect of processing on the Ti/Zr distribution in the thin films of industrially-relevant $PbTi_{1-x}Zr_xO_3$ whose functional properties have been reported as sensitive to the short-range B-cation order.[3]

## V. CONCLUSIONS

The effects of B-site short-range order and octahedral tilting on the pre-edge structure in the Ti-*K*-absorption spectra of perovskite solid solutions $CaTi_{1-x}Zr_xO_3$ were studied both experimentally and theoretically using full multiple-scattering calculations. The nature of the lowest unfilled conduction bands in this system was identified using the O *K*-XANES. The Ti *K*-spectra were calculated for a series of the 75-atom clusters $Ti(TiO_6)_{6-n}(ZrO_6)_nCa_8O_{24}$ with different concentrations of Zr ($p_{Zr}$) in the first B-cation coordination shell around the absorbing Ti atom (the effect of tilting was taken into account). The systematic differences in the pre-edge structure observed among the samples having different *x*-values were established to correlate closely with the B-cation short-range order parameter. These results demonstrate that Ti *K*-XANES can be used as a semi-quantitative tool for probing the degree of short-range B-cation order in the titanate perovskite-like solid solutions.

**ACKNOWLEDGEMENTS**

The authors are grateful to Dr. M. Lufaso for his help with the sample synthesis and to Dr. A. Novakovich for providing XKDQ code.

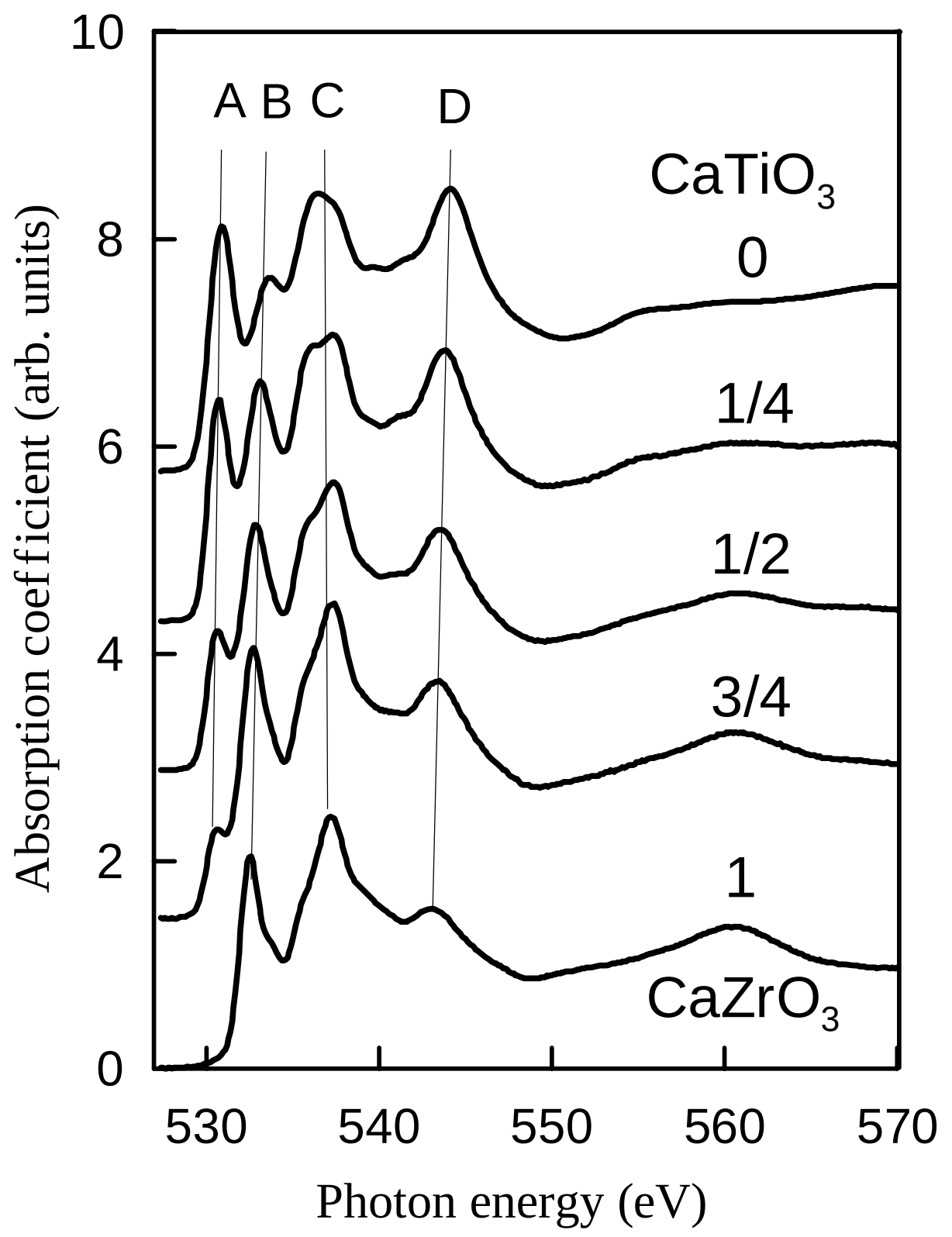

**FIG. 1. Oxygen *K*-XANES for $CaTi_{1-x}Zr_xO_3$ solid solutions with *x*=0, ¼, ½, ¾, and 1.**

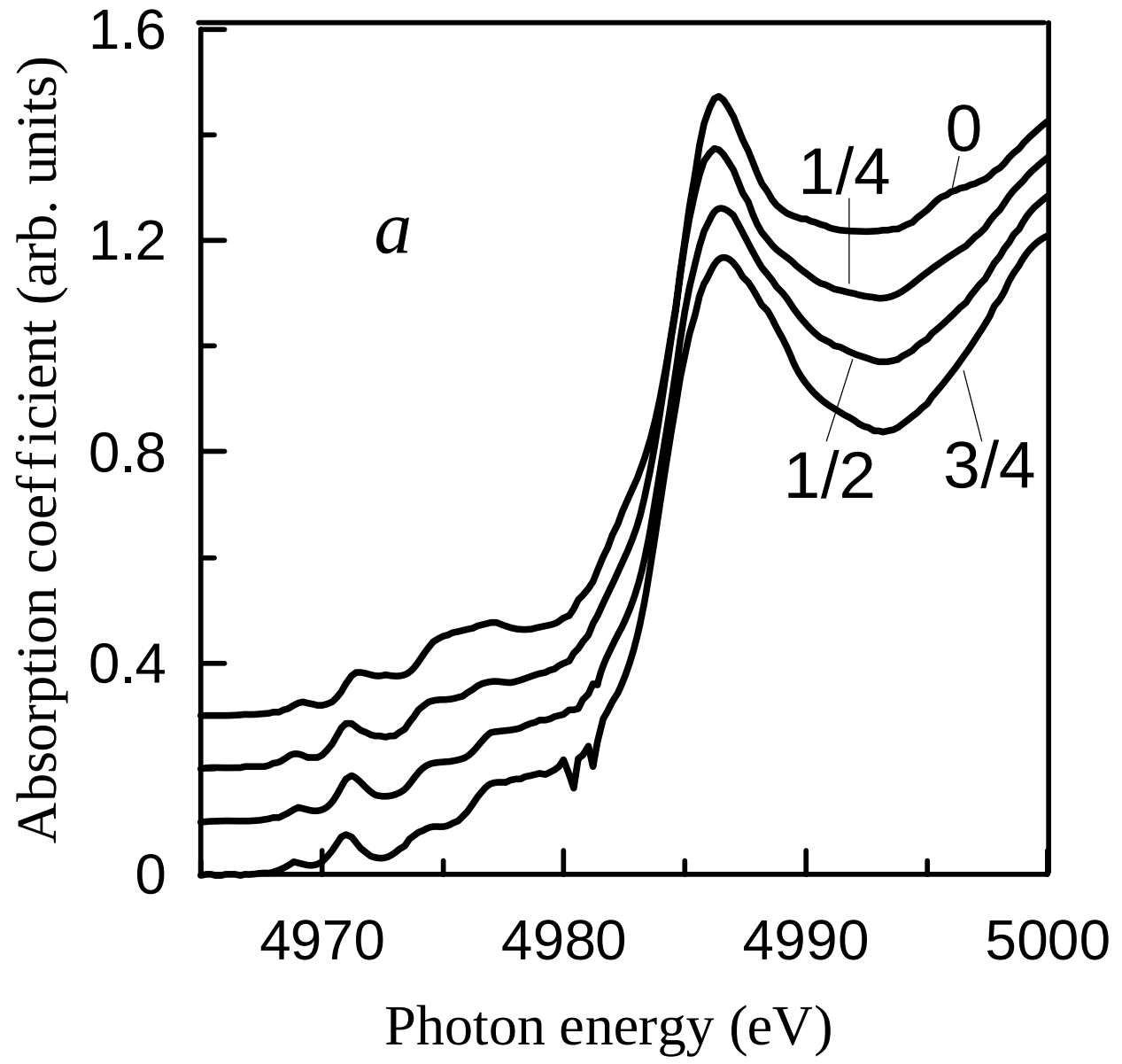

**FIG. 2. (a)-The Ti *K*-edge spectra of the $CaTi_{1-x}Zr_xO_3$ solid solutions with $x$=0, ¼, ½, and ¾. (b) - Enlarged pre-edge portions of the same spectra. The spectra were normalized by the intensities of the main edge peaks, and their pre-edge structures were aligned according to the positions of peak A and the low-energy side of peak B; the zero of the energy scale was chosen at the maximum of peak B.**

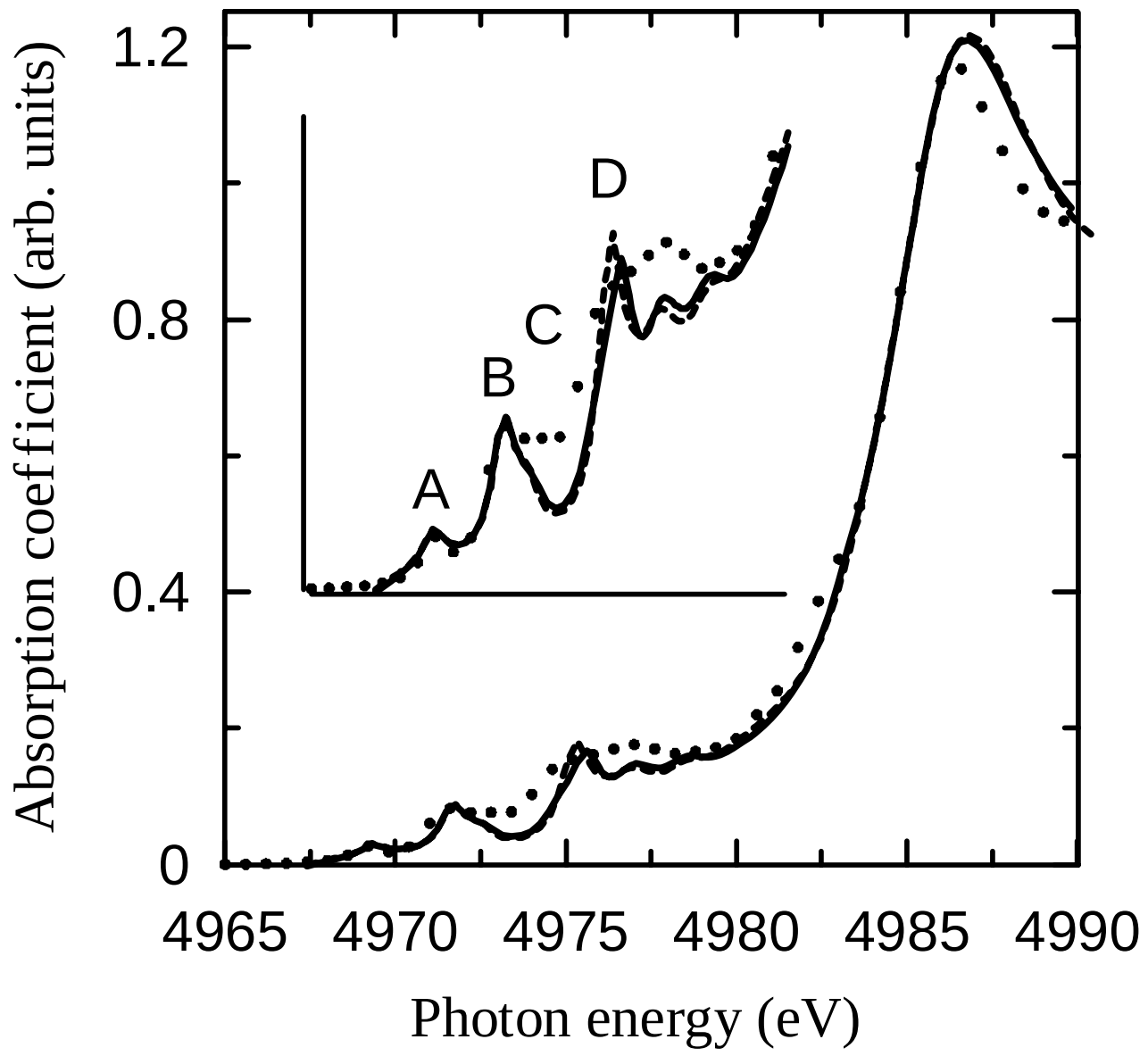

**FIG. 3. Experimental (dots) and calculated (line) Ti *K*-edge spectra for $CaTiO_3$. The theoretical spectra were computed for the 200- (solid line) and 75-atom (dashed line) clusters. All spectra are aligned according to the position and intensity of the main edge peak. The enlarged pre-edge part is shown in the insert.**

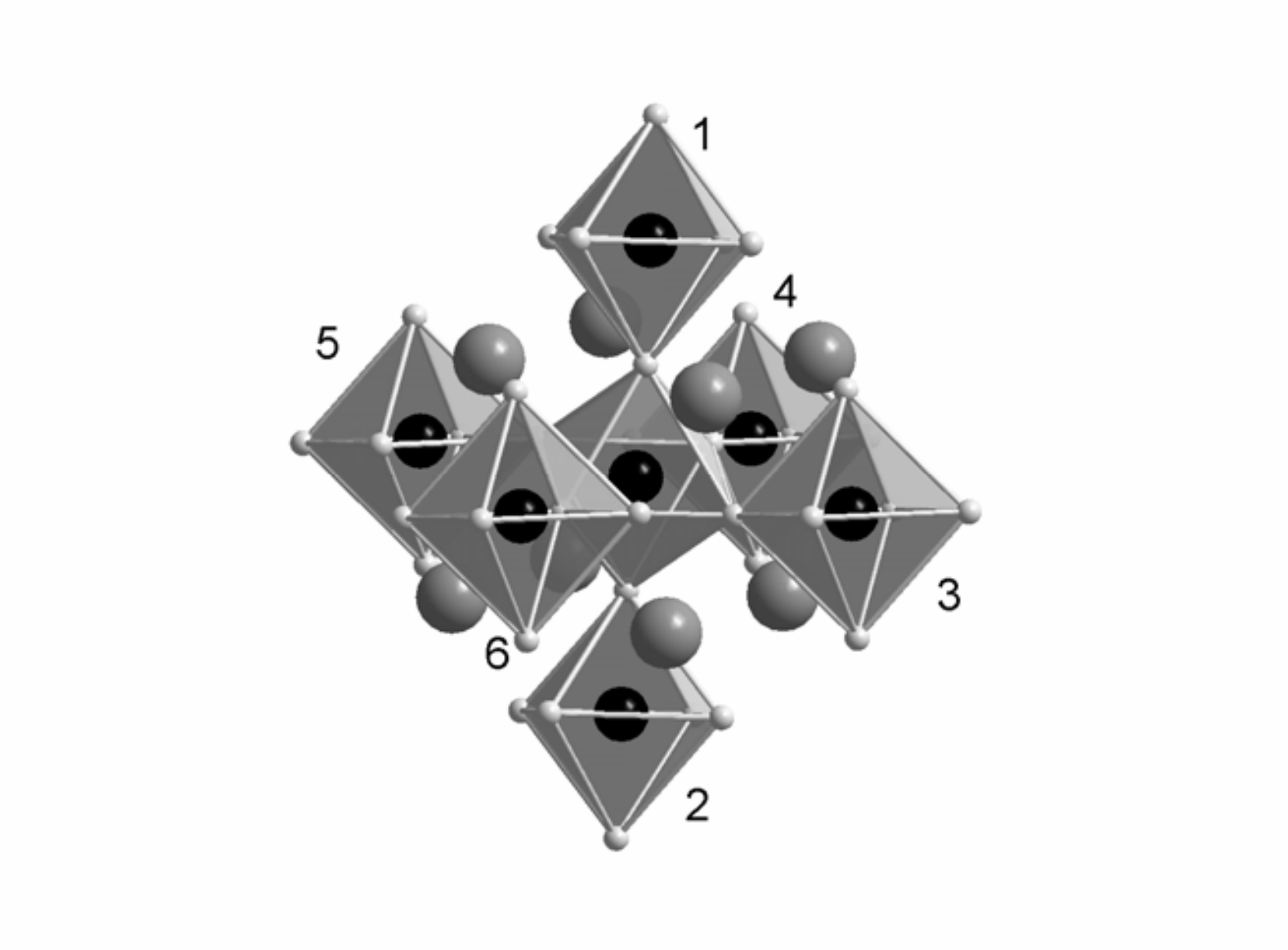

**FIG. 4. The 75–atom cluster of $CaTi_{1-x}Zr_xO_3$. Black spheres: Ti or Zr atoms, dark-gray spheres: Ca atoms, light-gray spheres: O atoms. 24 distant O atoms were omitted for clarity. The numbers specify location of specific $[TiO_6]$/$[ZrO_6]$ octahedra in the cluster.**

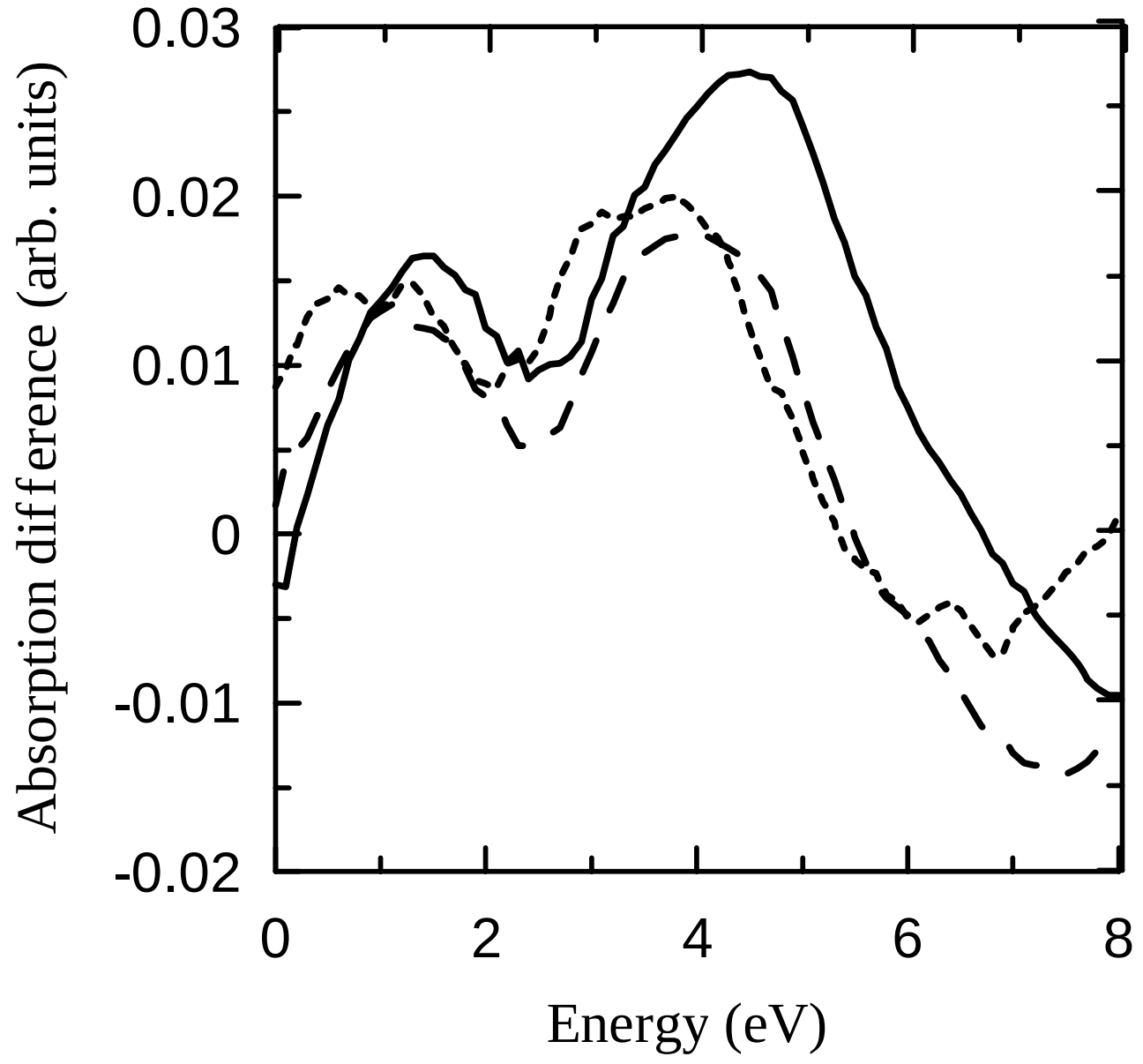

**FIG. 5. Experimental difference pre-edge structure obtained by subtracting the spectra of $CaTi_{¾-x}Zr_{¼+x}O_3$ from those of $CaTi_{1-x}Zr_xO_3$ for the following *x* values: 0-solid line, ¼-dashed line, and ½-dotted line.**

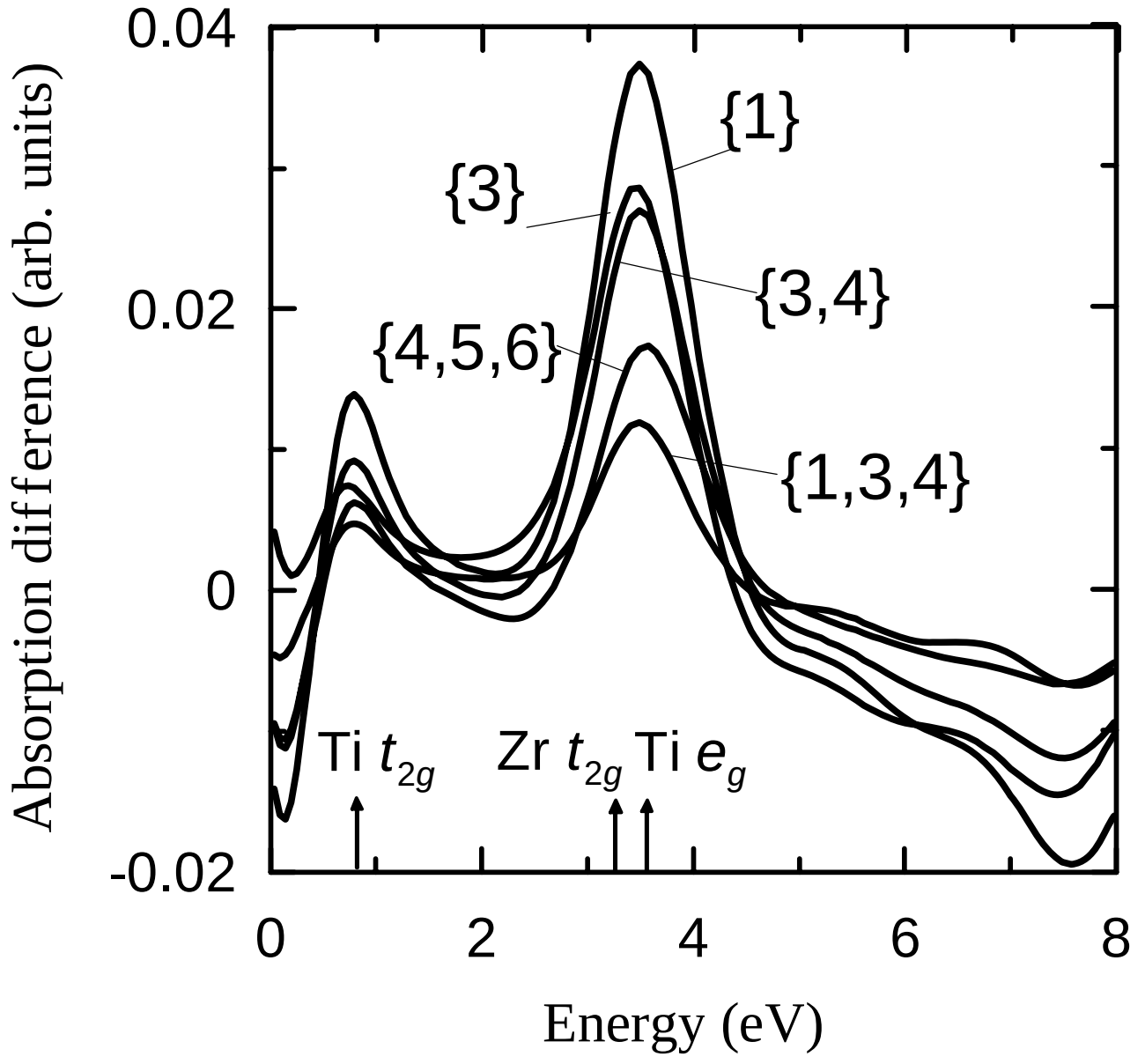

**FIG. 6. Theoretical difference pre-edge structure obtained by subtracting the spectra for the solid solution clusters $Ti(TiO_6)_{6-n}(ZrO_6)_nCa_8O_{24}$ (*n*=1, 2, 3) from that for the cluster $Ti(TiO_6)_6Ca_8O_{24}$. The following placements of $[ZrO_6]$ octahedra in the 75-atom cluster (see Fig. 4 for enumeration) were considered: {1}, {3}; {3, 4}; {4, 5, 6}, {1, 3, 4}. The arrows indicate the MO energies for the neighboring $[TiO_6]$/$[ZrO_6]$ octahedra.**

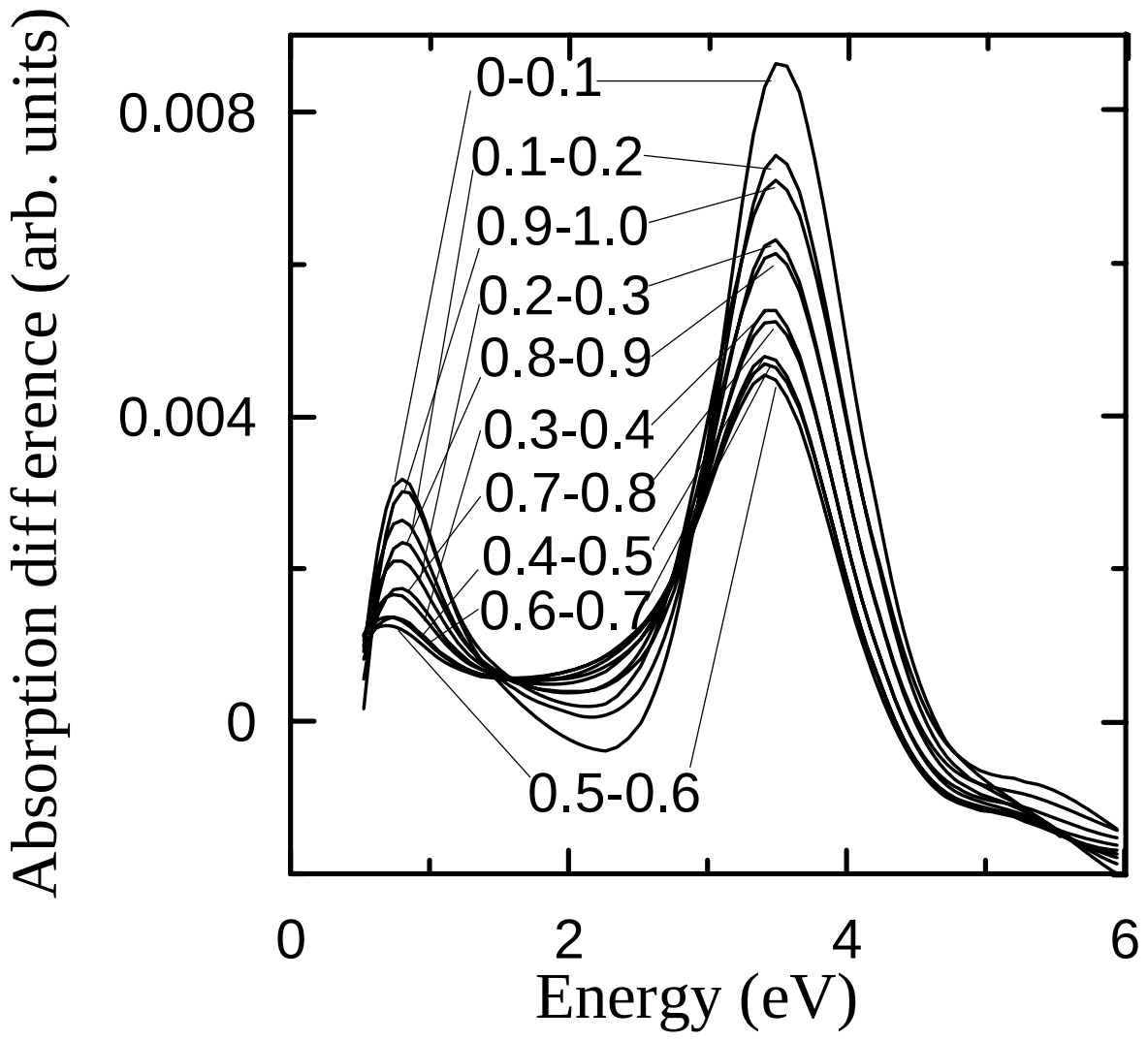

**FIG. 7. Theoretical difference pre-edge structure obtained according to $I_s(p_{Zr})$-$I_s(p_{Zr}+0.1)$ for $p_{Zr}$=0, 0.1, 0.2,…0.9.**

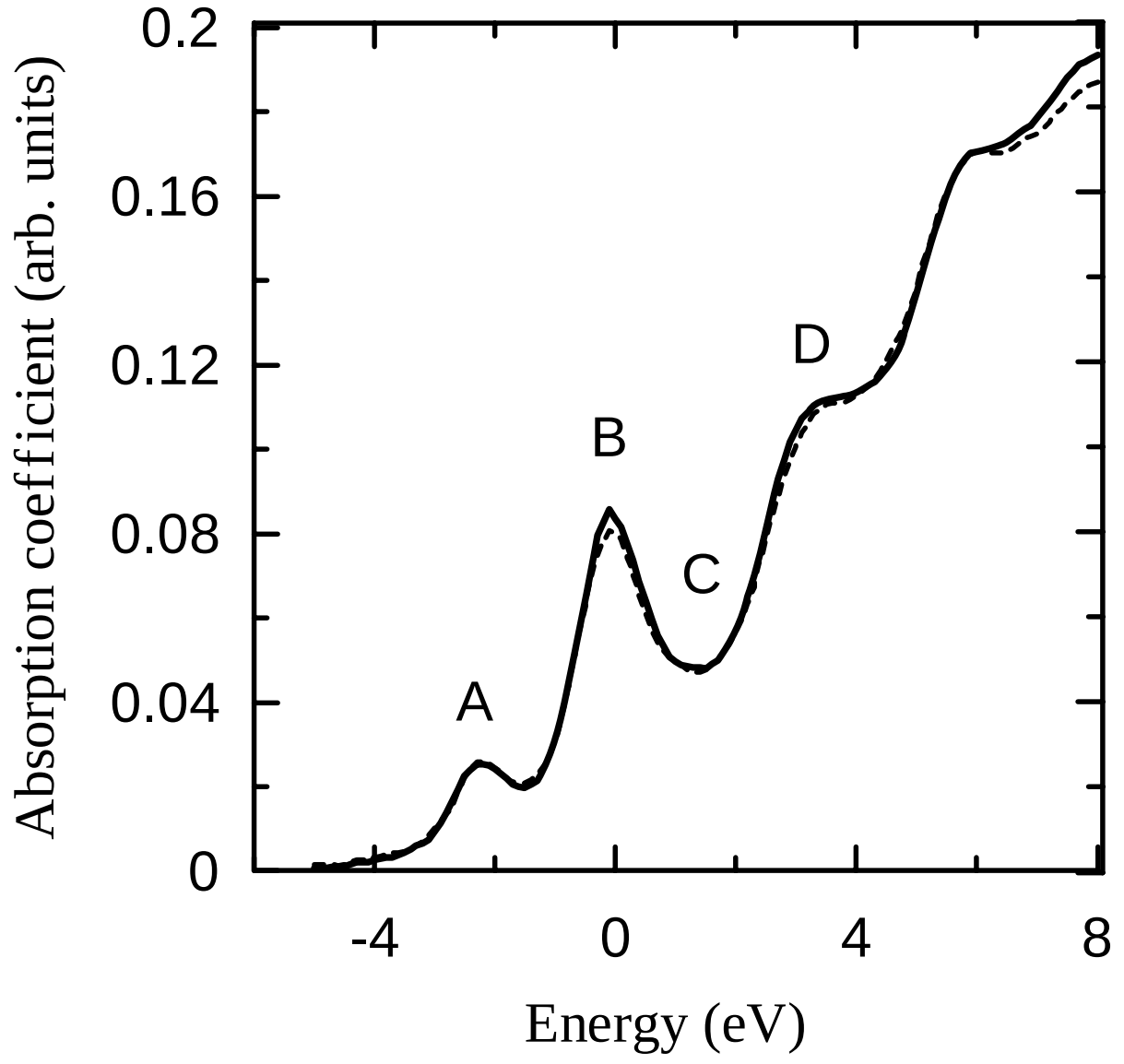

**FIG. 8. Experimental pre-edge structure for $CaTi_{1/2}Zr_{1/2}O_3$ (solid line) and the average of the experimental data for $CaTi_{1/4}Zr_{3/4}O_3$ and $CaTi_{3/4}Zr_{1/4}O_3$ (dashed line).**

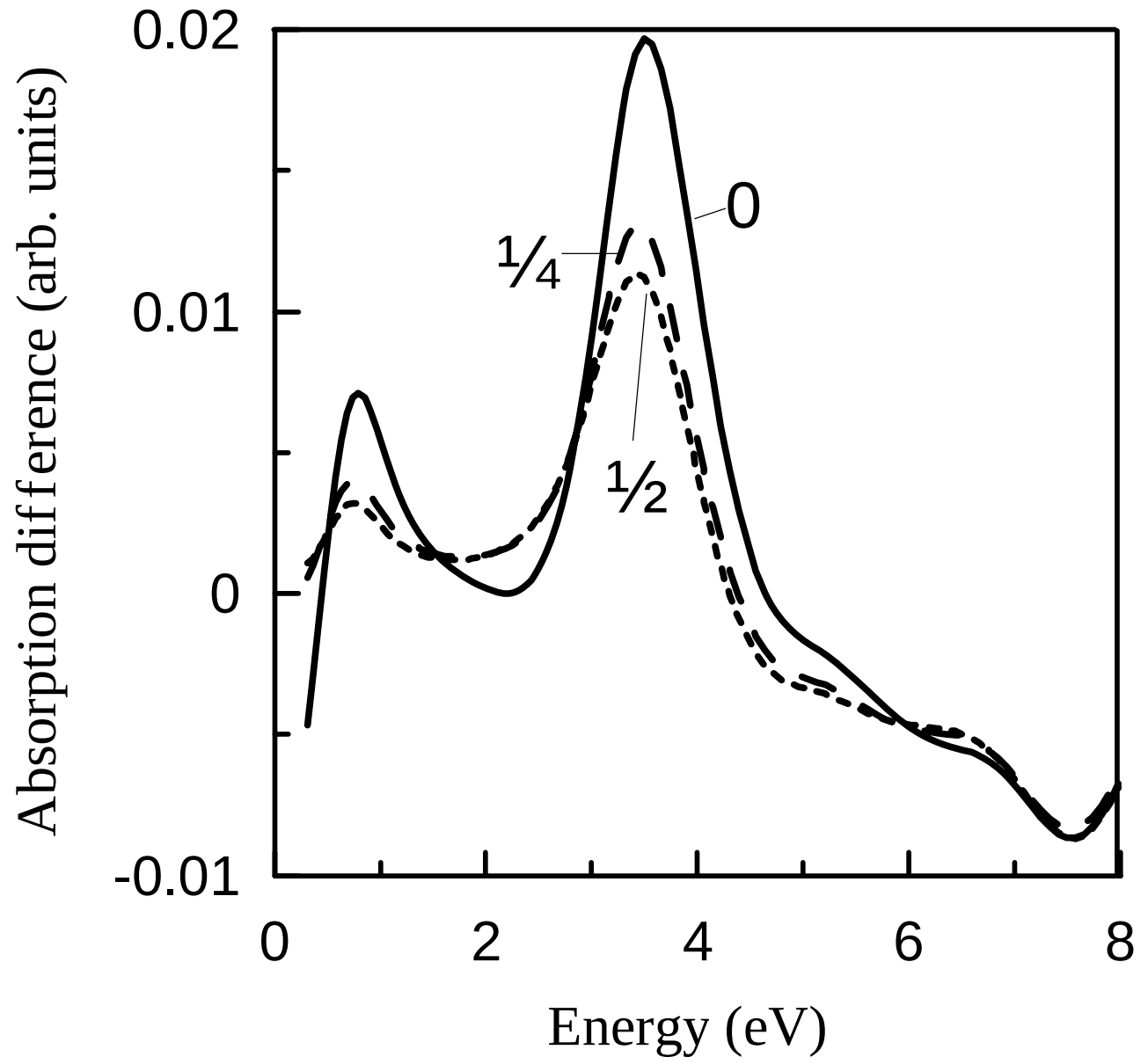

**FIG. 9. Theoretical difference spectra $I_s$ ($p_{Zr}$)-$I_s$ ($p_{Zr}$+¼) for $p_{Zr}$ =0 (solid line), ¼ (dashed line), and ½ (dotted line). Note a good qualitative agreement with the experimental difference spectra (see Fig. 5).**

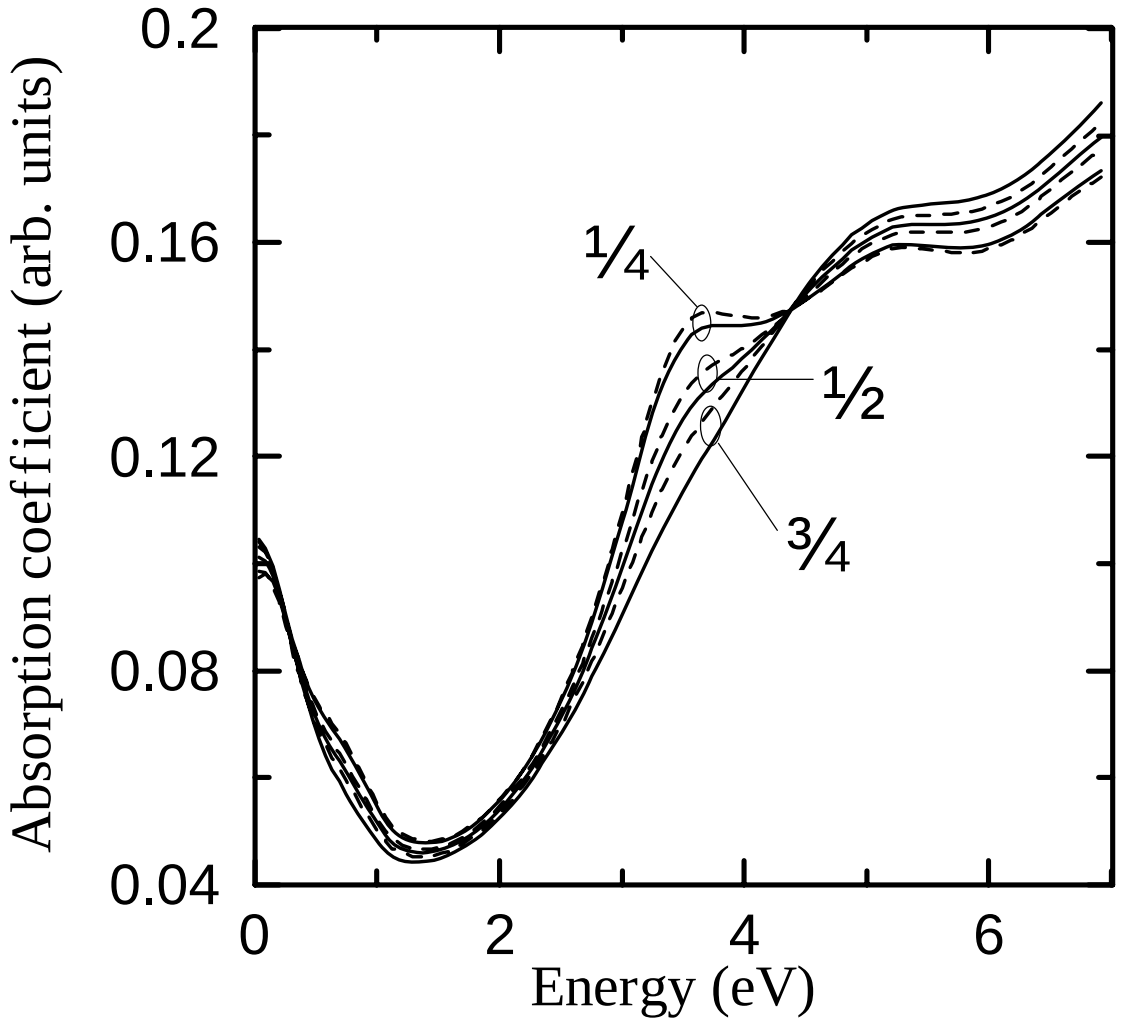

**FIG. 10. Ti *K*-pre-edge absorption calculated for the $CaTi_{1-x}Zr_xO_3$ solid solutions ($x$=¼, ½, and ¾) with the short-range order parameters $\vartheta$=0.1 (dashed lines) and $\vartheta$=-0.1 (solid lines). Note the effect of $\vartheta$ on the pre-edge structure for each composition.**

---

[21] K. Asokan, J.C. Jan, J.W. Chiou, W.F. Pong, M.-H. Tsai, H.L. Shih, H.Y. Chen, H.C. Hsueh, C.C. Chuang, Y.K. Chang, Y.Y. Chen, and I.N. Lin, J. Phys.: Condens. Matter **13**, 11087 (2001).

[22] T. Uozumi, K. Okada, A. Kotani, O. Durmeyer, J.P. Kappler, E. Beaurepaire, and J.C. Parlebas, Europhys. Lett. **18**, 85 (1992).

[23] J.M. Cowley, J. Appl. Phys. **21**, 24 (1950).

[24] M.W. Lufaso and P.M. Woodward Acta Crystallogr. B **57**, 725 (2001).